# A semi-automatic approach to study population dynamics based on population pyramids

Max Hahn-Klimroth[1*], João Pedro Meireles[2*], Laurie Bingaman Lackey[3], Nick van Eeuwijk[4] Mads F. Bertelsen[5], Paul W. Dierkes[1], Marcus Clauss[2]

[*]These authors contributed equally

[1]Goethe University Frankfurt, Frankfurt, Germany
[2]University of Zurich, Zurich, Switzerland
[3]1230 Oakland Street, Hendersonville, North Carolina, USA
[4]Utrecht University, Utrecht, The Netherlands
[5]Copenhagen Zoo, Frederiksberg, Denmark

Corresponding authors:

Max Hahn-Klimroth, Goethe University Frankfurt, Germany
Email: hahnklim@mathematik.uni-frankfurt.de, https://orcid.org/0000-0002-3995-419X

Marcus Clauss, Clinic for Zoo Animals, Exotic Pets and Wildlife, Vetsuisse Faculty, University of Zurich, Switzerland
Email: mclauss@vetclinics.uzh.ch, https://orcid.org/0000-0003-3841-6207

**Abstract**

The depiction of populations – of humans or animals – as 'population pyramids' is a useful tool for the assessment of various characteristics of populations at a glance. Although these visualisations are well-known objects in various communities, formalised and algorithmic approaches to gain information from these data are less present. Here, we present an algorithm-based classification of population data into 'pyramids' of different shapes ([normal and inverted] pyramid / plunger / bell, [lower / middle / upper] diamond, column, hourglass) that are linked to specific characteristics of the population. To develop the algorithmic approach, we used data describing global zoo populations of mammals from 1970-2024. This algorithm-based approach delivers plausible classifications, in particular with respect to changes in population size linked to specific series of, and transitions between, different 'pyramid' shapes. We believe this approach might become a useful tool for analysing and communicating historical population developments in multiple contexts and is of broad interest. Moreover, it might be useful for animal population management strategies.

# 1 Introduction

A standard demographic tool used to give a high-level description of a population is a so-called '*population pyramid*'. Here, typically, the male and female population sizes per age are used to give a visual snapshot of a population. Such a 'pyramid' contains a lot of demographic information, which can be interpreted and used to describe populations in different research fields: humanities, politics, ecology, as well as animal population management in zoological gardens. With improved and increased computational power and increasing data availability, options exist to not only view a single 'population pyramid' but also study sequences of 'pyramids' of the same population over time. However, it is challenging to analyse a sequence of graphical representations, and when comparing multiple populations, data normalisation methods are required. Moreover, depending on the resolution at which a population pyramid is expressed (i.e., the number of age classes), the data at hand is highly dimensional; therefore, meaningful analysis often comes with dimensionality reduction methods [1]. We propose a deterministic approach to reduce the dimensionality of the underlying population data, allowing us to compare different populations. Guided by the idea of classifying 'population pyramids' by their shape [2-4], we introduce a set of 'typical pyramid shapes' and a deterministic classification algorithm, which can be applied to a sequence of the modified 'population pyramids'. This enables us to study more abstract quantities regarding the evolution of the population by analysing sequences of 'pyramid' shapes. To test the plausibility of the classification approach, we compare demographic and biological interpretations regarding the shapes with the observable data. We introduce natural key measures such as a transition count between shapes, and we link shapes as well as transitions to typical properties of populations. The testing set contains over 50,000 'population pyramids' from more than 450 mammal species monitored in scientifically managed zoological institutions in Europe and North America. We believe that the approach itself can also be useful in other contexts where a series of 'population pyramids' needs to be studied. However, for the remainder of this contribution, we focus on animal population management.

## 1.1 Motivation

### 1.1.1 Monitoring Zoo Populations

Demographic analyses of zoo breeding programmes are essential for understanding small population dynamics, predicting future trends, and informing management and conservation strategies. Managed populations are monitored through the collection of studbook data [5], a detailed genetic and demographic record of a particular population. Managed either regionally or internationally, studbooks track the lineage and breeding history of individuals in the population. They guide pairing decisions to minimise inbreeding and to maximise genetic variation in order to maintain genetically viable populations. Furthermore, robust studbook data allows us to determine many other biological and demographic characteristics of the population, for instance, the typical age at first reproduction, birth sex allocation [6, 7], or survivorship dynamics [8, 9].

### 1.1.2 'Population Pyramids': A Common Display Method

A widely used graphical representation of population studbook data is the 'population pyramid', which visually depicts age and gender distribution [5, 10]. Here, we put the term into quotes because these 'pyramids' can have different shapes, only one of which actually resembles a pyramid (without quotes). These depictions consist of horizontal bars representing the size of different age cohorts, separated by gender [11]. The subset of the population that is non-reproductive (pre- or post-reproductive) is often highlighted. Depicting a 'population pyramid' is very helpful for understanding the sex and age distribution of a population, as this visual representation offers an easy and brief assessment, and also communication, of the breeding potential, the population growth, the population history, and potential present and future challenges. For instance, if a 'population pyramid' shows a declining juvenile population, it may signal reproductive issues, requiring targeted management and breeding techniques or husbandry improvements. In particular, the detection of demographic challenges benefits from an algorithmic approach towards such data rather than a simple visual inspection, i.e., if many populations over a longer period of time need to be monitored.

### 1.1.3 'Pyramids' Shaping Population Management

Small populations – as most zoo populations – are susceptible to stochastic events: a stroke of bad luck (e.g. an epidemic, transport restrictions, fires, political conflicts or unexpectedly high neonatal mortality) and the population may be jeopardised [5]. This risk becomes more significant the fewer individuals are actively reproducing in their reproductive prime, and the fewer juveniles are present for recruitment. Thus, zoo populations require intensive, careful management. Populations displaying a typical pyramid shape – a broad base and a narrow top – are more resilient to these stochastic events as the larger proportion of young, breeding individuals can buffer an occurrence with plenty of offspring. In contrast, a population with a narrow bottom (few young individuals) is at higher risk of suffering from negative events and hence less resilient (Figure 1).

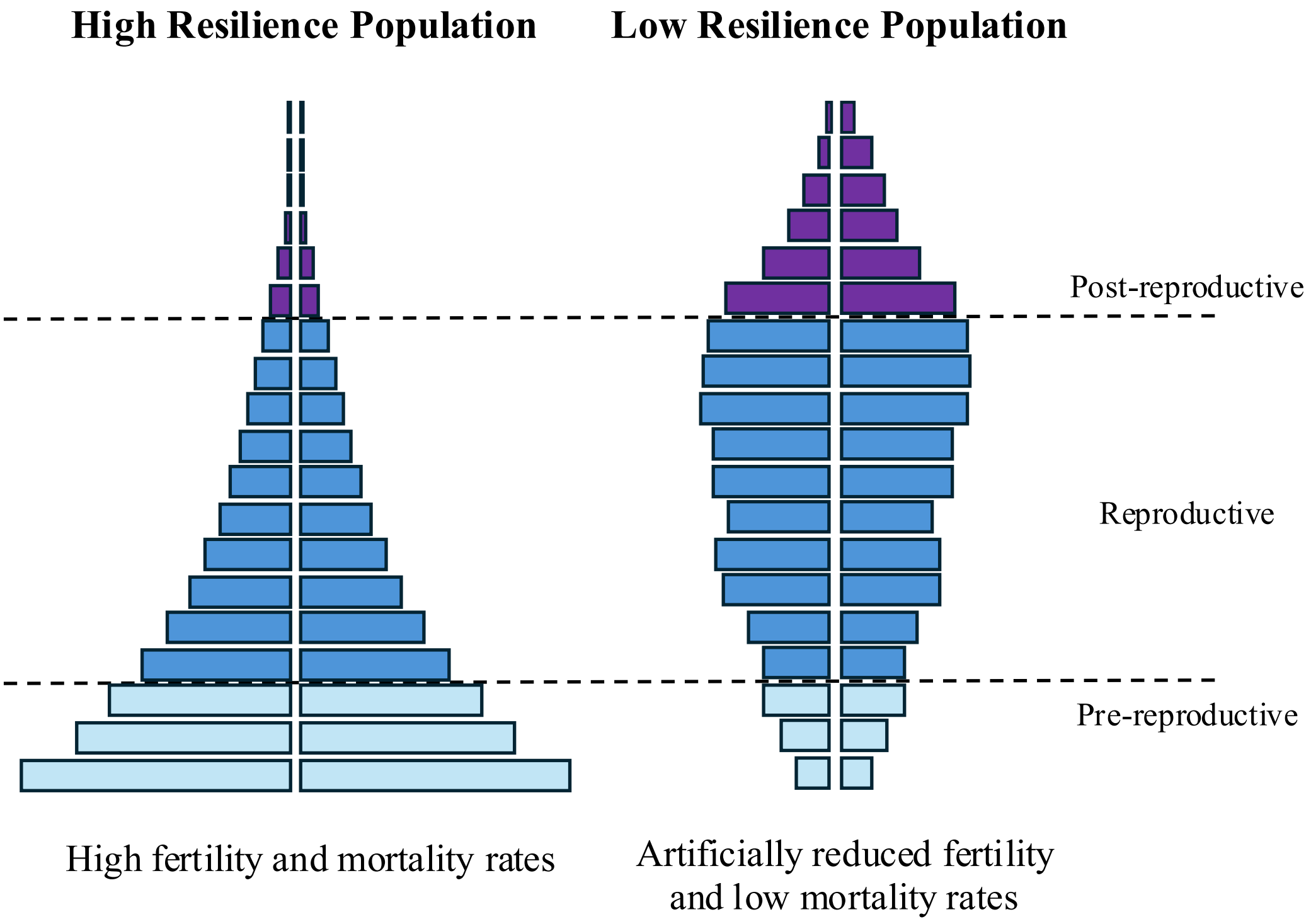

***Figure 1:*** *Population 'pyramids' showing the difference between a high-resilience population and a low-resilience population due to artificially reduced fertility and mortality.*

Therefore, dynamically analysing 'population pyramids' can help population managers and conservationists to make and communicate informed decisions regarding animal management to ensure long-term sustainability and viability of the breeding programmes [5, 10]. Decisions based on automatic detection of changes in the population pyramid shape can help to objectify decisions and to detect challenges earlier.

## 1.2 Contribution and related work

While 'population pyramids' are informative and intuitive, their interpretation can be subjective. In human demography, several authors have attempted to classify and aggregate 'population pyramids' to allow cross-population and over-time analyses. Korenjak-Černe et al. [2], Korenjak-Černe et al. [3] and Kosmelj and Billard [4] tried to cluster 'population pyramids' from different countries into homogeneous groups according to the similarity of their shape using different clustering methods. They successfully grouped countries that had similar ‘population pyramid’ shapes and could detect changes in countries between clusters, clearly following population structure trends.

Here, we aim to provide a deterministic method to automatically construct and classify ‘population pyramids’, rather than to rely on machine learning techniques. We defined a theoretically possible set of eleven distinct shapes (Table 1), predicting different probabilities of occurrence in real populations. We created a decision algorithm which first reduces the dimensionality of the ‘population pyramids’ and then classifies them in a deterministic and consistent manner, assigning one of the eleven theoretically possible shapes (Table 1). Similar to Ballou et al. [5], Rees [10] or Saroha [12], we additionally make predictions for the numerical population development for shape transitions (i.e., changes from one shape type to another) and a series of shapes (i.e., the same shape types occurring in subsequent years). Those predictions are used to test the accuracy and plausibility of the proposed classification algorithm. To assess these predictions, we count the number of different shapes that occur in mammal zoo populations and quantify the numerical population development associated with shape changes and shape series. The results suggest that the proposed algorithms can be used to reliably assess ‘population pyramid’ shapes.

Finally, as a prerequisite for drawing a population pyramid, one needs to define the ‘juvenile’, ‘adult’, and ‘senior’ life stages. As this is not well known for all species, we propose a method to estimate the corresponding age-group ranges based on a gamma distribution fitted to the probability that an individual reproduces at a given age.

# 2 Materials and methods

## 2.1 Source and selection of the data

The records of all individuals under the class Mammalia were obtained from Species360 (ZIMS for husbandry), an online database platform in current use by more than 1200 institutions worldwide to manage their animal data, under the licence agreement 103210. The dataset included information on the sex, dates of birth and death, whether the animal was wild or zoo-born, parentage data, geographic region in which it is held, and its current status (dead or alive) at the time of download. Birth dates for wild-born animals in the dataset were typically estimated at the time of import, and their reliability cannot be determined. Only species with at least 150 individuals ever recorded in Europe or North America were included. Next, our analysis focused on the European and North American populations of these species between January 1st 1970, to December 31st 2023, yielding a final number of 771 mammal populations. A total of 55,963 pyramids were automatically classified (771 × 2 (male and female pyramids) × ~54 years (not all populations existed since 1970)). This resulted in 54,081 transitions between pyramid shapes.

## 2.2 Data curation

Species360 data is prone to unreliable data points due to human error in the input of the data by its members. Hence, there was a need to curate the dataset extensively prior to its use. The following curation methodologies were pursued after the calculation of the age of the individual (as at 31 December 2023 or as at the time of death) and the age of each parent at the time of each birth event, both not part of the original data records:

- *Parentage*

  Only individuals with a 100% known probability of parentage (for both the dam or the sire) were considered as having parentage data. Among these, all those that had "impossible" parents, for instance, two parents of the same sex (the parent of incorrect sex is considered wrong), or being a parent of itself, had their parentage data removed (becoming parentless). In instances in which a male individual was marked as the dam and a female individual marked as the sire, the two were exchanged. Any individuals with parents younger than the defined age of first reproduction (see below) had their parent data removed.

- *Age of maximum lifespan*

  To define the maximum longevity for each species, the oldest zoo-born and dead individual was considered the record holder for the species. In some cases, the AnAge database [13] was consulted to assert those dubious cases when our stipulated rule would target exceptionally longevous individuals. Zoo-born individuals were chosen because the age of wild-born individuals is estimated, and one cannot be certain of their real age. Individuals currently alive at the top end of longevity could not be verified as reliable holders of such position (and often are the result of individuals forgotten to be marked as dead or lost-to-follow-up in ZIMS – thus immortal). Therefore, these cases were not considered for the age record of the species. It must be noted that this maximum longevity is not an attempt to determine the biologically possible maximum for the species, but merely the maximum observed in our dataset.

- *Interbirth intervals*

  To increase the likelihood of correct parentage data, interbirth intervals were calculated as the number of days between the birth of an individual and the birth of its immediate next sibling (as offspring of the same dam). Births occurring at less than 90% of the average gestation length were considered invalid (dam ID removed). For marsupials, due to inconsistent record methodology among institutions and difficulty in assessing births, the gestation length + minimum age of 1st pouch eviction was used instead. The minimum age of pouch eviction typical for each species was taken from the literature [14].

- *Litters*

  Individuals born from the same dam within 3 days were considered as part of the same litter, and their dates of birth were adjusted to a single (middle) date. In some 'nesting' species, it is common for zoo staff to only detect the presence of newborns of the same litter on different days. Litter sizes considered to be bigger than the known maximum for the species were considered as duplications and were cut in half by randomly removing half of their members.

# 3 Classification algorithm and 'pyramid' shapes

The main purpose of this contribution is the establishment of the aforementioned algorithmic classification of pyramids into different shapes and the analysis of the corresponding sequences.

## 3.1 A data-driven approach to define the senior life stage

To draw a typical 'population pyramid', one needs two age thresholds: the age of first reproduction ('the beginning of adult life stage') as well as the onset of reproductive senescence ('senior life stage'). While those thresholds are well studied for a few very prominent species and might be taken from the literature, for most species, the latter value might be particularly hard to access. From a management view, a 'typical' value indicating that reproductive activity becomes much lower after this age is of most interest; however, there may well be some older individuals still reproductively active.

We introduce a mathematical heuristic to determine this threshold. To calculate the *onset of reproductive senescence,* the proportion of proven breeders (individuals that have produced at least one offspring in their lifetime) reproducing in a given age class was calculated for each population. A gamma distribution was fitted to the sequence of age class probabilities. Then the age of post-peak-reproduction was defined as the age class at which 75% of the maximum of the gamma distribution was reached at the right tail (Figure 2). The use of a gamma distribution, which is often used to model natural processes rather than raw data, allows us to define a much less sensitive threshold to single outliers, a suggestion well known in the data science community, which nevertheless might be interesting on its own.

The age *of first reproduction* is much easier to estimate, as some observed births at this age are a good indication that individuals are reproductively active at this stage. Hence, the threshold was defined with the help of studbook data, bibliographic sources, or informed decisions based on species' biology or from members of the same Genus with more robust data/studbook. For both sexes, the age of first reproduction was taken as the time of the first birth to which they were associated as dams or sires.

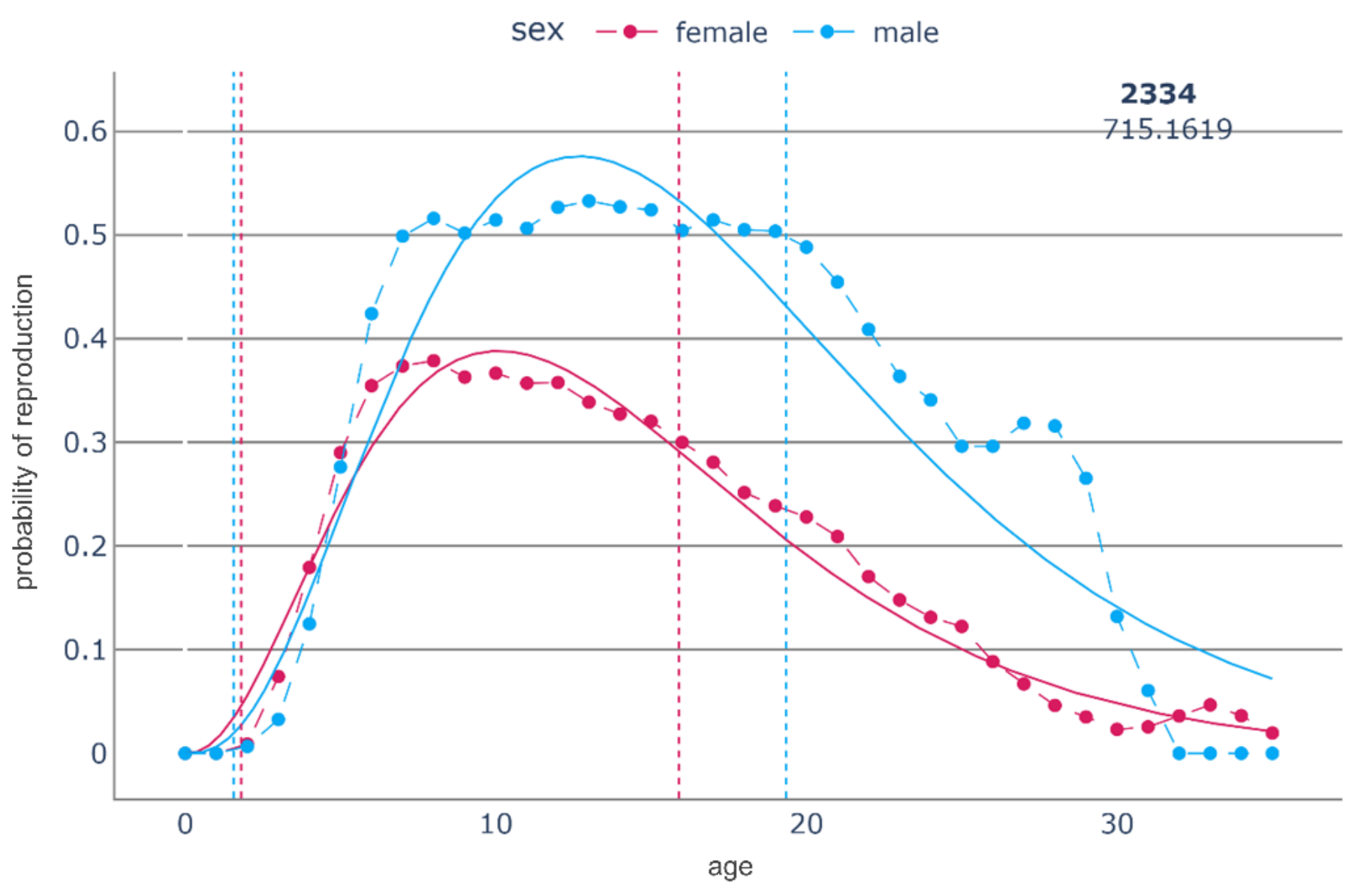

***Figure 2:*** *The probability of reproduction of the studied Plain's zebra (Equus quagga) population. The dots represent the empirical data while the lines represent the fitted gamma distributions. Vertical lines signify the age of first reproduction and the onset of reproductive senescence.*

## 3.2 'Pyramid' shape categories

'Population pyramids' were created for every year based on data from males and females separately. To assign individuals to age classes, their ages on December 31st of each year were calculated and floored to the next integer, e.g., an individual of age 363 days was said to have an age of 0 years (age class 0). For each age class, the number of individuals was counted and represented as a bar in the pyramid. These age classes were assigned to an age group (juvenile, adult, senior) based on the age of first reproduction and the age of onset of senescence, as described above.

To make an algorithmic approach to classify 'pyramids' of different species with different maximum ages, it is important to normalise the 'pyramids'. There are various ways of normalising data, and we decided to choose a method which represents a pyramid by five age buckets. Each 'pyramid' was divided into one block for juveniles, one for seniors, and three equal blocks for adults (Figure 3).

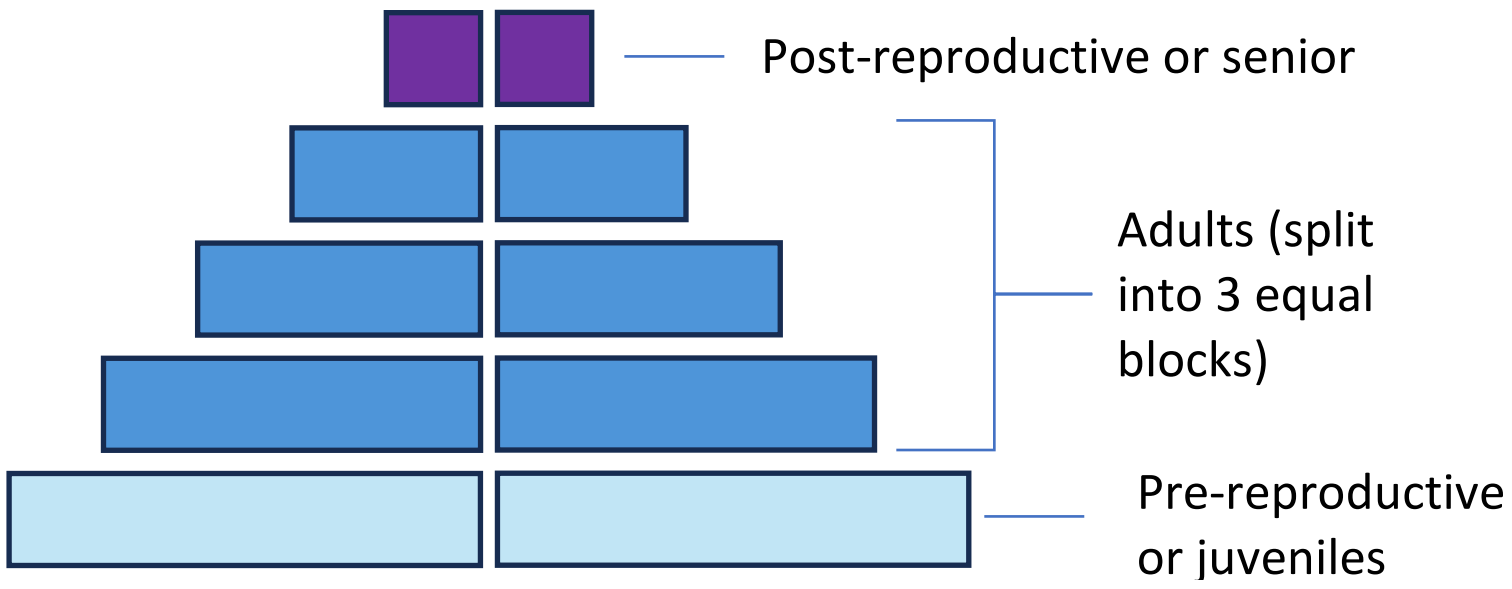

***Figure 3****: The division of an age 'pyramid' into 5 blocks.*

For each block, the mean number of individuals present per year in this block was assigned as the number of individuals. An equal number of individuals was assumed for two of those age blocks if those mean numbers ± 0.5 times their Standard Error of the Mean overlapped. This is not an over-conservative approach to detect differences, as this reflects that the 'typical real mean values are different'. As a data curation method, we propose that shapes are only ascribed if the sex-specific population contains more than 10 individuals, but this threshold might depend on the specific application.

Based on the five buckets, it is now possible to define distinct shapes (Table 1): (inverted) pyramid, (inverted) bell, (inverted) plunger, diamond (depending on the broadest part, divided into lower, middle and upper diamond), column, and hourglass.

***Table 1:*** *The proposed classification of population 'pyramids' into different shapes based on the reduced pyramid variant. The shapes can be distinguished by a sequence of length four, describing 'increase', 'stationary', 'decrease' when analysing the transition from the n-th bucket to the (n+1)st bucket. This mapping is not 1:1, as there are many transition sequences that need to be classified differently depending on the relative size of non-consecutive buckets, see Table S8 in the appendix for a complete mapping.*

| Pyramid Types | Description |
|---|---|
| **Pyramid/Inverted Pyramid** 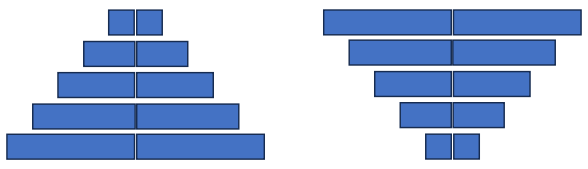 | Continuous increase/decrease |
| **Plunger/Inverted Plunger** 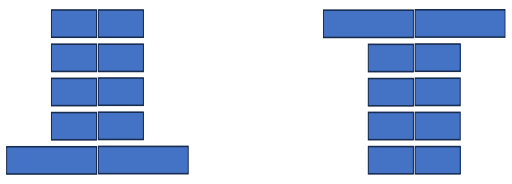 | Long, thin unchanging part and a broader part |
| **Bell/Inverted Bell** 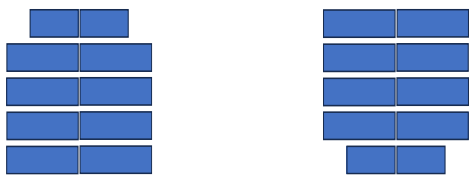 | Long, broad unchanging part and a thinner part |
| **Diamond (middle, lower and upper variants)** 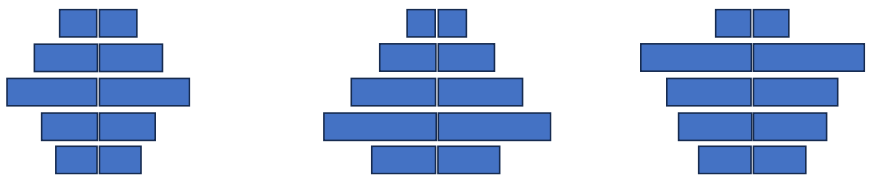 | The broader part is surrounded by thinner parts, depending on location divided into middle, lower, upper |
| **Column** 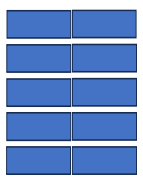 | No change |
| **Hourglass** 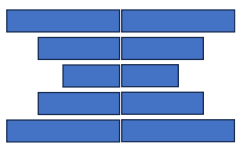 | Thinner part surrounded by broader parts |

## 3.3 Algorithmic detection of the 'pyramid' shape

Using the described normalisation method, we can compare pyramids of different species and sizes, but we also managed to reduce data complexity dramatically. Indeed, given a reduced pyramid, we can compare the bucket sizes and get a sequence of four steps with the values +1, 0, -1 representing 'becoming larger', 'having the same size', and 'becoming smaller' which describes the transition from one bucket to another (starting from the lowest bucket). Here, 'having the same size' is meant as described before (when the mean numbers ± 0.5 times their Standard Error of the Mean overlapped).

Unfortunately, not all shapes look as perfect as the typical shape classes, as is usual in any data classification task. This means that, for instance, the sequence of four steps is not enough to describe all cases well enough, but for some sequences, we need to compare more buckets. For example, the sequence (-1, -1, 1, 1) can describe a pyramid or an hourglass, depending on the size difference of the first and the third 'adult bucket'. We suggest a comprehensive classification rule, given in the supplementary material.

# 4 Assessing the plausibility of the automated shape categorisation

## 4.1 Frequency of shape transitions

Having determined the 'pyramid' shape, we can now introduce demographic key figures. A standard measure is the *transition frequency* from one shape to another. There are several natural assumptions about how we would imagine typical transition frequencies, and we evaluated those assumptions against the zoo population data. The most common shape transition by far was from pyramid to pyramid, i.e. a sequence of pyramids. Generally, series or sequences of the same shape (the diagonal of Table 2) were more common than transitions changing to a different shape. Among changes of shapes (from one shape type to a different one), the most frequently observed was from pyramid to lower diamond. Changes from pyramid to bell, bell to pyramid and lower diamond to pyramid were also common, followed by the change from lower to middle and from middle to upper diamond (Table 2).

***Table 2:*** *Percentage of occurrence of shape transitions (total n=54081). Only cases with percentages above 0.1% are displayed. Cases with percentages above 1% are shaded in grey. Cases above 2% are marked in bold.*

| From/to | 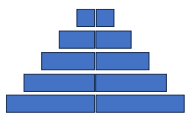 | 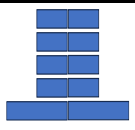 | 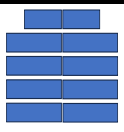 | 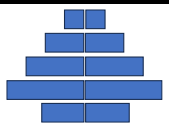 | 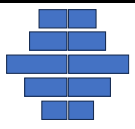 | 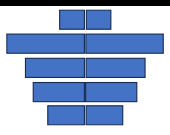 | 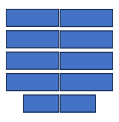 | 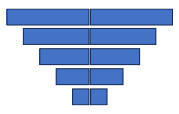 | 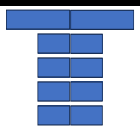 | 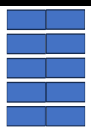 | 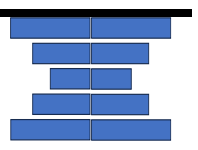 |
|---|---|---|---|---|---|---|---|---|---|---|---|
| 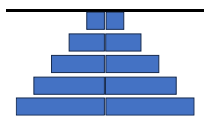 | **50.3%** | 1.0% | **3.4%** | **5.7%** | 0.8% | 0.2% | - | - | - | - | 0.4% |
| 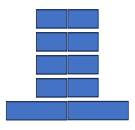 | 1.2% | 0.8% | 0.1% | 0.4% | - | - | - | - | - | - | 0.2% |
| 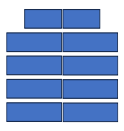 | **3.8%** | 0.2% | **3.1%** | 1.2% | 1.1% | 0.6% | - | - | - | 0.1% | 0.2% |
| 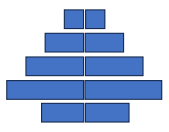 | **3.2%** | 0.1% | 1.8% | **4.2%** | **2.0%** | 0.3% | - | - | - | - | 0.1% |
| 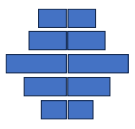 | 0.9% | - | 1.2% | 0.4% | **2.1%** | 1.2% | - | - | - | - | 0.1% |
| 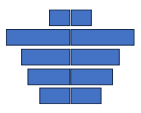 | 0.4% | 0.1% | 0.6% | 0.1% | 0.3% | **2.2%** | 0.1% | - | - | - | 0.3% |
| 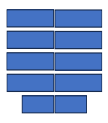 | - | - | - | - | - | - | 0.1% | - | - | - | - |
| 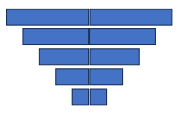 | - | - | - | - | - | - | - | 0.1% | - | - | - |
| 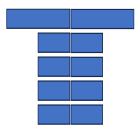 | - | - | - | - | - | - | - | - | 0.1% | - | - |
| 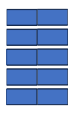 | - | - | 0.1% | - | - | - | - | - | - | - | - |
| 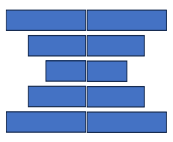 | 0.5% | 0.2% | 0.1% | 0.3% | 0.1% | 0.1% | - | - | - | - | 0.6% |

As a condensed variant of Table 2, we give a comparison between our expectations on the transition frequency and the real occurrences in the dataset in Table 3. Most expectations were met, which indicates that the classification algorithm seems to work well on average.

***Table 3:*** *Prediction of frequency of occurrence of shape transitions (white fields) or series of the same shapes (grey fields); green symbols represent predictions supported by the data, whereas orange symbols represent unexpected results.*

| From/to | 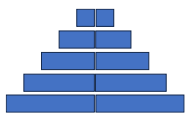 | 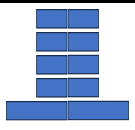 | 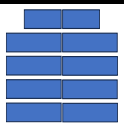 | 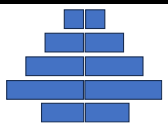 | 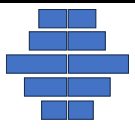 | 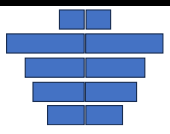 | 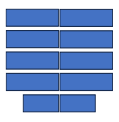 | 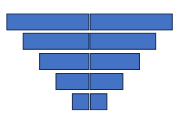 | 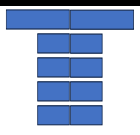 | 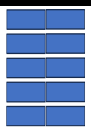 | 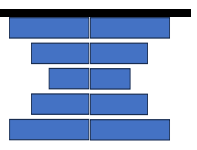 |
|---|---|---|---|---|---|---|---|---|---|---|---|
| 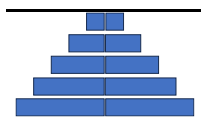 | +++ | - | + | + | - | - | -- | --- | --- | -- | - |
| 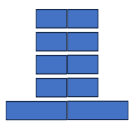 | + | - | - | - | - | - | -- | --- | --- | - | --- |
| 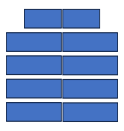 | + | - | ++ | + | + | - | -- | --- | --- | + | - |
| 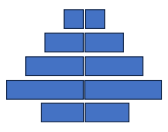 | + | - | + | + | + | + | - | --- | --- | - | -- |
| 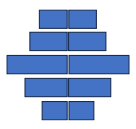 | --- | -- | -- | -- | + | + | - | - | - | - | -- |
| 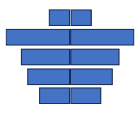 | --- | --- | --- | --- | -- | + | -- | - | - | - | - |
| 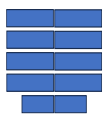 | --- | -- | -- | - | - | - | - | - | -- | - | - |
| 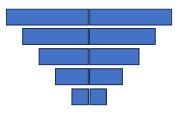 | --- | --- | --- | --- | --- | - | - | - | - | -- | - |
| 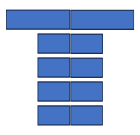 | --- | - | -- | --- | --- | --- | --- | --- | -- | - | -- |
| 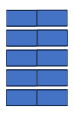 | -- | + | + | -- | -- | -- | -- | --- | --- | + | -- |
| 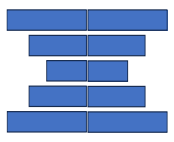 | + | --- | --- | - | -- | -- | -- | --- | --- | - | - |

## 4.2 Evolution of the population size with shape transitions

A single ‘pyramid’ shape already contains much information about the population: a prediction of *the development of the population size*. Given a pyramid, we would, for instance, suppose that the overall population size is increasing while an inverted pyramid occurs in decreasing populations. While Table 5 contains the most important population size differences, our expectations and the comparison to the data are given in Table 6. Generally, changes from any shape to a pyramid or a plunger were associated with a population increase, whereas changes to diamonds and columns were associated with a population decrease (Table 5). Decreases were especially pronounced in changes to columns. Transitions to bells had the lowest average population size change, with decreases when coming from pyramids or plungers, increases when coming from diamonds, and a transition from bell to bell had a negligible net change in population size (Table 5). Again, the predicted directions of population size changes with shape transitions were largely met, with the prominent exceptions that transitions to plungers were more often associated with a population increase than expected (Table 6).

***Table 5****: Average (±SD if n ≥ 2) population change of shape-to-shape transitions in %. Changes are only displayed when the transition frequency is above 0.1%. Positive population changes are shaded in blue and negative in orange. In bold, the largest population increase and decrease. The average of each column summarises the displayed values for changes to a certain shape.*

| From/to | | | | | | | | | | | |
|---|---|---|---|---|---|---|---|---|---|---|---|
| | 7.2%<br>(±8.5) | 5.3%<br>(±15.9) | -2.9%<br>(±8.4) | -4.0%<br>(±8.2) | -6.8%<br>(±10.7) | -6.0%<br>(±9.4) | - | - | - | - | 2.5%<br>(±17.7) |
| | 8.8%<br>(±19.3) | 6.2%<br>(±16.0) | -4.4%<br>(±11.2) | -4.1%<br>(±11.7) | - | - | - | - | - | - | 3.7%<br>(±23.7) |
| | 7.1%<br>(±12.9) | 3.0%<br>(±12.1) | 0.1%<br>(±8.3) | -4.1%<br>(±8.3) | -4.0%<br>(±7.2) | -5.2%<br>(±8.7) | - | - | - | -2.0%<br>(±9.7) | -0.1%<br>(±11.9) |
| | **9.5%**<br>(±16.3) | 7.6%<br>(±14.5) | 2.2%<br>(±9.9) | -3.4%<br>(±7.1) | -3.9%<br>(±7.6) | -5.7%<br>(±10.0) | - | - | - | - | 3.6%<br>(±18.2) |
| | 7.2%<br>(±13.3) | - | 2.1%<br>(±10.1) | -4.3%<br>(±7.7) | -3.9%<br>(±7.3) | -4.9%<br>(±8.1) | - | - | - | - | -0.6%<br>(±12.4) |
| | 4.4%<br>(±14.0) | 2.3%<br>(±14.1) | 0.9%<br>(±11.1) | -7.6%<br>(±11.5) | -5.2%<br>(±8.3) | -4.3%<br>(±7.1) | -6.7%<br>(±7.0) | - | - | - | -0.9%<br>(±9.1) |
| | - | - | - | - | - | - | -9.1%<br>(±5.9) | - | - | - | - |
| | - | - | - | - | - | - | - | -8.9%<br>(±6.0) | - | - | - |
| | - | - | - | - | - | - | - | - | -3.4%<br>(±6.5) | - | - |
| | - | - | -1.2%<br>(±7.3) | - | - | - | - | - | - | - | - |
| | 8.0%<br>(±18.0) | 4.1%<br>(±11.6) | -0.9%<br>(±9.9) | -5.6%<br>(±12.2) | **-10.8%**<br>(±9.4) | -6.4%<br>(±9.7) | - | - | - | - | 4.0%<br>(±17.5) |
| **Average** | 7.4%<br>(±1.5) | 4.8%<br>(±1.8) | -0.5%<br>(±2.2) | -4.7%<br>(±1.3) | -5.8%<br>(±2.5) | -5.4%<br>(±0.7) | -7.9%<br>(±1.2) | -8.9% | -3.4% | -2% | 1.7%<br>(±2.0) |

***Table 6:*** *Predictions of population changes between shape transitions (white fields) or series of the same shapes (grey fields); predictions supported by the data in green, those not supported by the data in purple; those that occurred at very low frequency without colours*

| From/to | 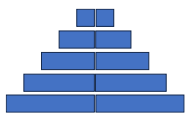 | 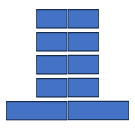 | 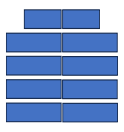 | 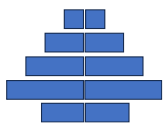 | 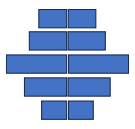 | 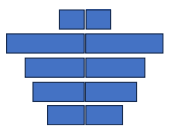 | 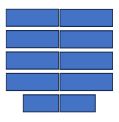 | 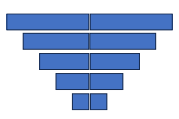 | 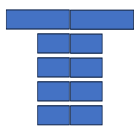 | 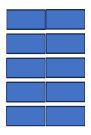 | 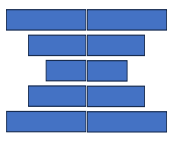 |
|---|---|---|---|---|---|---|---|---|---|---|---|
| 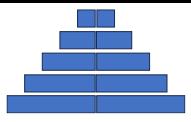 | ∧ | ∨ | ∨ | ∨ | ∨ | ∨ | - | - | - | - | ∨ |
| 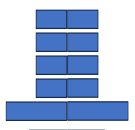 | ∧ | = | ∧ | ∨ | ∨ | - | - | - | - | ∨ | ∧ |
| 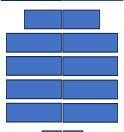 | ∧ | ∧ | = | ∨ | ∨ | ∨ | - | - | - | ∨ | = |
| 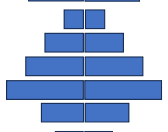 | ∧ | ∧ | ∧ | ∨ | ∨ | ∨ | ∧ | - | - | ∨ | = |
| 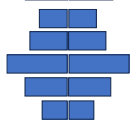 | ∧ | = | ∧ | ∨ | ∨ | ∨ | ∧ | - | - | ∨ | = |
| 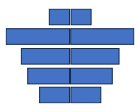 | ∧ | = | ∧ | ∨ | ∨ | ∨ | ∨ | ∨ | ∨ | ∨ | ∧ |
| 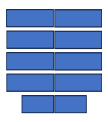 | ∧ | ∧ | = | ∨ | ∨ | ∨ | ∨ | ∨ | ∨ | ∨ | ∧ |
| 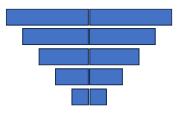 | - | - | - | ∨ | - | - | ∨ | ∨ | ∨ | ∨ | ∧ |
| 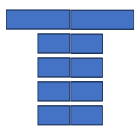 | - | = | ∨ | - | ∧ | ∧ | ∧ | ∧ | ∨ | ∨ | ∧ |
| 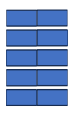 | ∧ | ∧ | ∨ | ∧ | ∧ | ∧ | ∧ | ∧ | ∧ | = | ∧ |
| 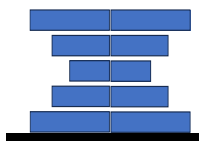 | = | ∧ | ∨ | ∨ | ∨ | ∨ | - | ∨ | ∨ | ∨ | ∨ |

## 4.3 Proportion of age groups

Each 'pyramid' shape comes with a typical distribution of the proportion of juveniles, adults, and seniors. The proportions of age groups for each shape follow what is expected. Shapes like pyramid, plunger or bell have a larger proportion of juveniles and a lower proportion of seniors, while shapes like inverted pyramid, inverted plunger or inverted bell have a larger proportion of senior animals (Table 7).

***Table 7:*** *Percentage of age groups in different population shapes. Maximum and minimum observed proportions for each age group are shaded in orange and blue, respectively.*

| | 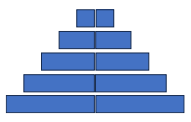 | 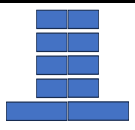 | 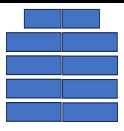 | 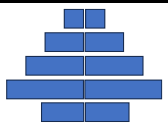 | 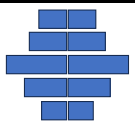 | 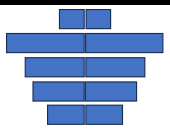 | 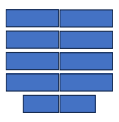 | 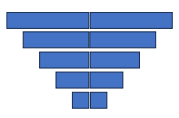 | 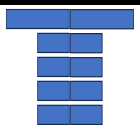 | 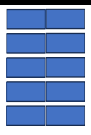 | 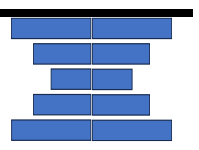 |
|---|---|---|---|---|---|---|---|---|---|---|---|
| **Seniors** | 10.7% (±10.3) | 22.2% (±18.7) | 21.1% (±14.8) | 18.3% (±14.5) | 25.3% (±15.3) | 36.4% (±17.3) | 62.0% (±14.5) | 76.9% (±13.3) | 67.1% (±23.8) | 54.9% (±15.3) | 36.8% (±22.8) |
| **Adults** | 60.5% (±12.3) | 44.3% (±16.0) | 60.8% (±14.1) | 70.4% (±13.9) | 65.5% (±14.5) | 56.1% (±16.5) | 34.7% (±13.2) | 22.2% (±13.0) | 14.5% (±12.6) | 33.8% (±13.2) | 38.9% (±18.1) |
| **Juveniles** | 28.7% (±12.9) | 33.5% (±20.8) | 18.1% (±9.3) | 11.2% (±7.5) | 9.2% (±7.2) | 7.6% (±6.9) | 3.3% (±4.2) | 0.9% (±2.0) | 18.4% (±26.3) | 11.2% (±7.4) | 24.3% (±18.7) |

## 4.4 Evolution of the population size over time

We evaluate the shape at maximum population size, and for each other shape, calculate the average population size (as % of the maximum population size for the species) of that shape. In our dataset, pyramid shapes were by far the most frequent, followed by bell and lower, middle and upper diamonds (in that sequence). Plunger, column and hourglass were at similar, low frequencies, whereas the unlikely inverted pyramid, inverted plunger and inverted bell occurred very rarely. The vast majority of the (female and male) populations reached their peak size with a population structure of a pyramid shape. Pyramids, bells, lower and middle diamonds were generally associated with population sizes closer to the peak, followed by upper diamonds and hourglasses; by contrast, columns were associated with low population sizes (Table 8), in line with natural expectations.

***Table 8:*** *Count of occurrences of pyramid shapes and the associated population sizes*

| | 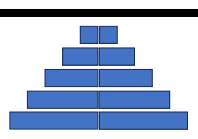 | 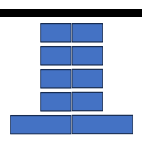 | 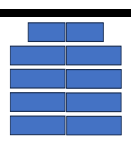 | 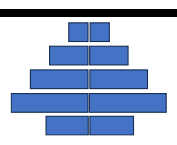 | 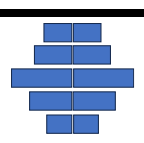 | 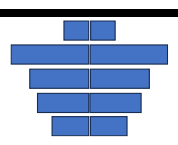 | 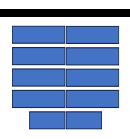 | 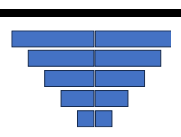 | 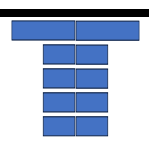 | 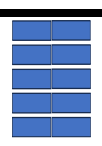 | 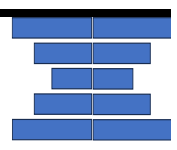 |
|---|---|---|---|---|---|---|---|---|---|---|---|
| **Total occurrences** | 33960 | 1550 | 5832 | 6739 | 3495 | 2540 | 193 | 104 | 152 | 177 | 1221 |
| **% of occurrence** | 60.7% | 2.8% | 10.4% | 12.0% | 6.2% | 4.5% | 0.3% | 0.2% | 0.3% | 0.3% | 2.2% |
| **Occurrence of shapes at maximum pop. size** | 1114 | 47 | 144 | 137 | 42 | 32 | 0 | 0 | 1 | 1 | 30 |
| **Average percentage of the maximum pop. size between 1970-2023 per shape** | 66.2% (11.6) | 53.9% (22.5) | 64.6% (19.0) | 65.2% (18.1) | 64.2% (18.9) | 61.4% (19.6) | 47.5% (21.4) | 48.8% (21.9) | 41.1% (18.9) | 48.8% (19.4) | 53.0% (21.9) |

# 5 Discussion

## 5.1 Technical strengths and limitations

Probably the most obvious limitation of the proposed approach is that the 11 possible shapes and thus the decision rules by which they were assigned were defined ‘arbitrarily’. While they do have some biological, natural interpretations, other possible shapes could be thought of. However, as the classification problem was reduced to an assignment based on five buckets, the shapes seem to be well chosen. A second limitation lies in this same data simplification. Although reducing the input dimension in classification problems is a standard way of working in data science and machine learning [15-17], this automatically comes with a loss of information, and more severely, with an inapplicability to certain datasets. In our case, one prerequisite for a species is that the ‘adult range’ consists of at least three years, as otherwise the age buckets are not well defined. In such cases, one can, of course, for instance, work with ‘half-years’ instead of ‘years’, but this requires adaptation of the algorithm. However, we believe that a ‘one-size-fits-all ’ approach is highly unlikely to exist in any data application.

We observed that most expectations of the studied key figures were met by the classification rules. This indicates that the rules, as well as the choice of the shapes, seem to be in line with natural assumptions. As the underlying dataset is quite large, consisting of 55,963 pyramids, we believe that our observations are valid and the classification rules, as well as the data simplification approach, work well. Finally, a second strength of the proposed method is that the approach is completely comprehensive and deterministic. It does not depend on complicated machine learning algorithms or a dataset-wise classification rule for the pyramid shapes, as other classification approaches do [3], which makes it more reliable in different applications.

## 5.2 Content-wise interpretation

In this contribution, we introduced an algorithmic approach to study population pyramids, their shapes, and sequences of such shapes. The algorithm was tested and tuned on a large dataset of zoo animal populations, which is why we will briefly discuss some observations content-wise in this topic. However, we strongly believe that the approach can also be used in other demographic studies, such as within social sciences, and is of broader interest. Moreover, our focus is not on the content-wise interpretation but on the description as well as promotion of the methodology of how to deal with such data.

Of the theoretically possible eleven shapes, three biologically unlikely ones were expectedly very rare among zoo mammal populations – the inverted pyramids, inverted bells and inverted plungers. These shapes occur if the oldest age class shows either an increase or no decline compared to the adult stage, which can be considered a rare fortuitous event, as actuarial senescence, i.e. an increase in mortality with age, is a common characteristic of mammals and birds [18-21]. Therefore, the shape types of pyramid, bell and diamond appear the most important, and most frequent ones, and their upper parts always reflect the reality of senescence, with a decreasing number of senior individuals.

A pyramid shape means that a large number of offspring is produced relative to the number of adults, and that not all of these make it to the adult stage. A bell indicates that there is little mortality between juveniles and adults, which can be considered rarer compared to distinct juvenile mortality [21]. A diamond represents a situation in which juvenile production was reduced compared to previous years, and the duration of the time of reduced juvenile production determines whether the broadest part of the diamond is on its lower, middle or upper part. The characteristics of these shapes are aptly reflected in the population size changes associated with shape transitions and shape series, which mostly correspond to the predictions made by us and others [5, 10].

The information visually concentrated in ‘population pyramids’ can also be gleaned from tabulated data. In our view, the main advantage of using ‘population pyramids’ is the efficacy of communication, because the visual patterns are intuitively accessible. In discussions about the state of populations and decisions about population management, they can inform larger audiences effectively [11]. This, we believe, will be particularly important in assessing historical trends to not only understand the past, but to make informed decisions for the future of zoo population management. The method outlined in the present manuscript can be applied to such evaluations in the future.

# 6 Statements

## 6.1 Data availability

The code used to conduct all calculations in the Python programming language is available on our GitHub repository: https://github.com/Klimroth/population_pyramids/
The original contributions presented in the study are included in the article/appendix. Further inquiries can be directed to the corresponding author.

## 6.2 Ethics statement

Not applicable.

## 6.3 Competing interests

The authors declare no conflict of interest.

## 6.4 Consent to publish

Not applicable.

## 6.5 Consent to participate

Not applicable.

## 6.6 Funding

MHK and PWD were supported by Opel-Zoo Foundation Professorship in Zoo Biology from the "von Opel Hessische Zoostiftung". JPM was supported by Zoo Zurich.

## 6.7 Author contribution

Conceptualisation: MHK, JPM, LBL, MFB, PWD, MC
Methodology: MHK, JPM, MC
Visualisation: MHK, JPM
Formal analysis: MHK, JPM, LBL, MC
Data curation: JPM, LBL, NvE, MC
Funding acquisition: PD, MC
Supervision: MFB, PWD, MC
All authors contributed to the writing of the original draft and agreed upon submission.

## 6.8 Acknowledgements

This research was made possible by the worldwide information network of zoos and aquariums which are members of Species360 and is authorized by Species360 Research Data Use Agreement # 103210. We thank Severin Dressen from Zoo Zurich for his support of this study.

***Supplementary Material***

# A method for the automated evaluation of zoo animal population pyramids

Max Hahn-Klimroth, João Pedro Meireles, Laurie Bingaman Lackey, Nick van Eeuwijk, Mads F. Bertelsen, Paul W. Dierkes, Marcus Clauss

***Table S1:*** *Frequency (counts) of shape series.*

| Length (years) | 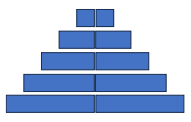 | 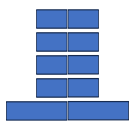 | 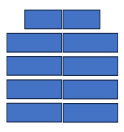 | 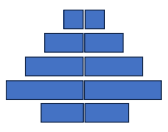 | 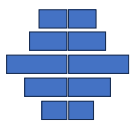 | 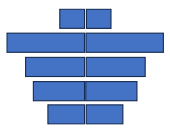 | 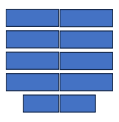 | 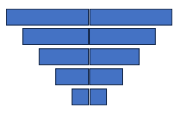 | 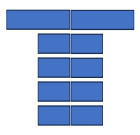 | 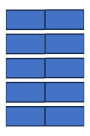 | 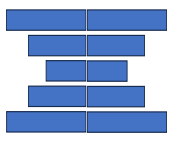 |
|---|---|---|---|---|---|---|---|---|---|---|---|
| **2** | 1184 | 212 | 775 | 1040 | 521 | 264 | 19 | 6 | 20 | 18 | 149 |
| **3** | 780 | 51 | 237 | 328 | 167 | 136 | 7 | 11 | 4 | 2 | 40 |
| **4** | 551 | 18 | 85 | 123 | 54 | 59 | 1 | - | 2 | 1 | 18 |
| **5** | 365 | 6 | 30 | 34 | 10 | 32 | - | - | 1 | - | 7 |
| **6** | 276 | 5 | 6 | 8 | 9 | 22 | - | - | 1 | - | 1 |
| **7** | 192 | 1 | 2 | 6 | 2 | 9 | - | - | - | - | 2 |
| **8** | 144 | - | 1 | - | - | 8 | - | 1 | - | - | - |
| **9** | 125 | 1 | - | - | - | 5 | - | - | - | - | - |
| **10** | 80 | - | - | - | - | 3 | - | - | - | - | 1 |
| **11** | 91 | 1 | - | - | - | 2 | - | - | - | - | 1 |
| **12** | 71 | - | - | - | - | - | - | - | - | - | - |
| **13** | 62 | - | - | - | - | 2 | - | - | - | - | - |
| **14** | 47 | - | - | - | - | 1 | - | - | - | - | - |
| **15** | 39 | - | - | - | - | - | - | - | - | - | - |
| **16** | 37 | - | - | - | - | - | - | - | - | - | - |
| **17** | 29 | - | - | - | - | - | - | - | - | - | - |
| **18** | 25 | - | - | - | - | - | - | - | - | - | - |
| **19** | 22 | - | - | - | - | - | - | - | - | - | - |
| **20** | 22 | - | - | - | - | - | - | - | - | - | - |

***Table S2:*** Average (±SD if n ≥ 2) population change from start to end of shape series.

| Length (years) | 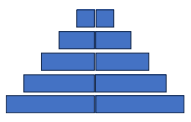 | 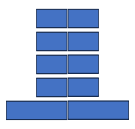 | 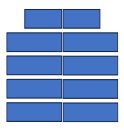 | 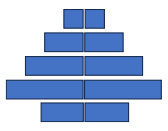 | 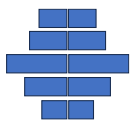 | 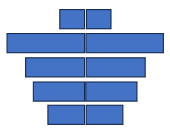 | 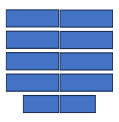 | 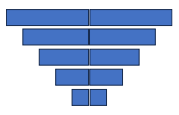 | 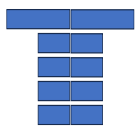 | 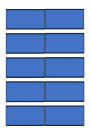 | 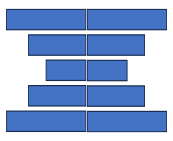 |
|---|---|---|---|---|---|---|---|---|---|---|---|
| **2** | 4.4% (±14.5) | 5.3% (±16.3) | -0.1% (±9.4) | -3.7% (±8.1) | -3.8% (±8.1) | -4.2% (±9.0) | -10.5% (±6.9) | -8.1% (±6.5) | -3.1% (±6.4) | -1.8% (±6.2) | 4.0% (±19.7) |
| **3** | 12.3% (±23.0) | 13.9% (±26.4) | 0.2% (±12.5) | -5.6% (±10.8) | -7.6% (±11.3) | -8.2% (±11.9) | -14.5% (±9.1) | -17.2% (±11.4) | -11.3% (±9.2) | -3.9% (±3.9) | 8.8% (±21.7) |
| **4** | 24.0% (±46.4) | 24.0% (±63.6) | -1.0% (±15.4) | -6.7% (±12.4) | -10.6% (±13.3) | -13.8% (±12.5) | -5.6% | - | -11.2% (±7.4) | -31.2% | 7.0% (±37.9) |
| **5** | 28.3% (±43.0) | 67.0% (±95.4) | 5.2% (±25.9) | -6.9% (±11.3) | -8.0% (±18.8) | -17.8% (±13.7) | - | - | -38.9% | - | 30.4% (±49.9) |
| **6** | 42.5% (±53.1) | 27.5% (±41.5) | -6.2% (±18.4) | 3.3% (±8.6) | -10.1% (±12.0) | -11.3% (±17.5) | - | - | 50.0% | - | -27.4% |
| **7** | 49.2% (±64.1) | 23.9% | 36.4% (±40.0) | 7.3% (±13.4) | -2.9% (±10.5) | -20.4% (±14.7) | - | - | - | - | -44.7% (±7.5) |
| **8** | 49.0% (±70.6) | - | 36.4% | - | - | -24.8% (±14.4) | - | -64.0% | - | - | - |
| **9** | 74.8% (±87.9) | 84.6% | - | - | - | -40.0% (±20.2) | - | - | - | - | - |
| **10** | 78.1% (±86.9) | - | - | - | - | -15.2% (±12.4) | - | - | - | - | -48.3% |
| **11** | 91.2% (±103.4) | 190.0% | - | - | - | -28.2% (±14.0) | - | - | - | - | -14.6% |
| **12** | 123.5% (±133.9) | - | - | - | - | - | - | - | - | - | - |
| **13** | 119.0% (±144.0) | - | - | - | - | -17.7% (±2.9) | - | - | - | - | - |
| **14** | 125.3% (±104.6) | - | - | - | - | -41.0% | - | - | - | - | - |
| **15** | 127.8% (±135.8) | - | - | - | - | - | - | - | - | - | - |
| **16** | 140.9% (±164.5) | - | - | - | - | - | - | - | - | - | - |
| **17** | 169.1% (±176.8) | - | - | - | - | - | - | - | - | - | - |
| **18** | 295.3% (±311.2) | - | - | - | - | - | - | - | - | - | - |
| **19** | 134.4% (±136.8) | - | - | - | - | - | - | - | - | - | - |
| **20** | 311.9% (±393.3) | - | - | - | - | - | - | - | - | - | - |

**Table S3** Number of shape transitions (total n=54081).

| From/to | pyramid | plunger | bell | lower diamond | middle diamond | upper diamond | inverted bell | inverted pyramid | inverted plunger | column | hourglass |
|---|---|---|---|---|---|---|---|---|---|---|---|
| **pyramid** | 27198 | 521 | 1860 | 3066 | 420 | 114 | 3 | 1 | 9 | 11 | 243 |
| **plunger** | 631 | 441 | 70 | 206 | 13 | 5 | 4 | 0 | 3 | 17 | 105 |
| **bell** | 2075 | 122 | 1673 | 625 | 597 | 314 | 18 | 6 | 6 | 36 | 126 |
| **lower diamond** | 1719 | 52 | 982 | 2277 | 1078 | 167 | 23 | 3 | 1 | 13 | 45 |
| **middle diamond** | 495 | 23 | 646 | 212 | 1114 | 654 | 26 | 4 | 3 | 12 | 41 |
| **upper diamond** | 231 | 44 | 347 | 59 | 187 | 1185 | 38 | 25 | 24 | 18 | 155 |
| **inverted bell** | 5 | 6 | 10 | 18 | 21 | 15 | 36 | 9 | 7 | 14 | 13 |
| **inverted pyramid** | 0 | 1 | 1 | 2 | 3 | 4 | 12 | 35 | 13 | 3 | 12 |
| **inverted plunger** | 12 | 7 | 7 | 6 | 2 | 5 | 5 | 8 | 43 | 4 | 21 |
| **column** | 16 | 17 | 32 | 19 | 10 | 6 | 9 | 3 | 7 | 25 | 12 |
| **hourglass** | 265 | 125 | 69 | 153 | 28 | 62 | 19 | 10 | 23 | 22 | 347 |

**Table S4** Percentage of shape transitions (total n=54081).

| From/to | pyramid | plunger | bell | lower diamond | middle diamond | upper diamond | inverted bell | inverted pyramid | inverted plunger | column | hourglass |
|---|---|---|---|---|---|---|---|---|---|---|---|
| **pyramid** | 50.3% | 1.0% | 3.4% | 5.7% | 0.8% | 0.2% | 0.0% | 0.0% | 0.0% | 0.0% | 0.4% |
| **plunger** | 1.2% | 0.8% | 0.1% | 0.4% | 0.0% | 0.0% | 0.0% | 0.0% | 0.0% | 0.0% | 0.2% |
| **bell** | 3.8% | 0.2% | 3.1% | 1.2% | 1.1% | 0.6% | 0.0% | 0.0% | 0.0% | 0.1% | 0.2% |
| **lower diamond** | 3.2% | 0.1% | 1.8% | 4.2% | 2.0% | 0.3% | 0.0% | 0.0% | 0.0% | 0.0% | 0.1% |
| **middle diamond** | 0.9% | 0.0% | 1.2% | 0.4% | 2.1% | 1.2% | 0.0% | 0.0% | 0.0% | 0.0% | 0.1% |
| **upper diamond** | 0.4% | 0.1% | 0.6% | 0.1% | 0.3% | 2.2% | 0.1% | 0.0% | 0.0% | 0.0% | 0.3% |
| **inverted bell** | 0.0% | 0.0% | 0.0% | 0.0% | 0.0% | 0.0% | 0.1% | 0.0% | 0.0% | 0.0% | 0.0% |
| **inverted pyramid** | 0.0% | 0.0% | 0.0% | 0.0% | 0.0% | 0.0% | 0.0% | 0.1% | 0.0% | 0.0% | 0.0% |
| **inverted plunger** | 0.0% | 0.0% | 0.0% | 0.0% | 0.0% | 0.0% | 0.0% | 0.0% | 0.1% | 0.0% | 0.0% |
| **column** | 0.0% | 0.0% | 0.1% | 0.0% | 0.0% | 0.0% | 0.0% | 0.0% | 0.0% | 0.0% | 0.0% |
| **hourglass** | 0.5% | 0.2% | 0.1% | 0.3% | 0.1% | 0.1% | 0.0% | 0.0% | 0.0% | 0.0% | 0.6% |

**Table S5** Average population change per shape transition.

| From/to | pyramid | plunger | bell | lower diamond | middle diamond | upper diamond | inverted bell | inverted pyramid | inverted plunger | column | hourglass |
|---|---|---|---|---|---|---|---|---|---|---|---|
| pyramid | 7.2% | 5.3% | -2.9% | -4.0% | -6.8% | -6.0% | -12.0% | -7.7% | -3.5% | -6.3% | 2.5% |
| plunger | 8.8% | 6.2% | -4.4% | -4.1% | -5.4% | -12.2% | -8.1% | - | -9.4% | -5.3% | 3.7% |
| bell | 7.1% | 3.0% | 0.1% | -4.1% | -4.0% | -5.2% | -5.4% | -7.0% | -14.5% | -2.0% | -0.1% |
| lower diamond | 9.5% | 7.6% | 2.2% | -3.4% | -3.9% | -5.7% | -5.8% | -9.1% | -9.1% | -4.1% | 3.6% |
| middle diamond | 7.2% | 4.5% | 2.1% | -4.3% | -3.9% | -4.9% | -8.1% | -14.8% | -9.3% | -2.4% | -0.6% |
| upper diamond | 4.4% | 2.3% | 0.9% | -7.6% | -5.2% | -4.3% | -6.7% | -9.2% | -9.1% | -8.2% | -0.9% |
| inverted bell | 14.2% | 7.1% | -1.8% | -6.5% | -10.1% | -8.8% | -9.1% | -3.8% | -1.8% | -0.6% | -1.1% |
| inverted pyramid | - | -17.6% | 5.6% | -1.5% | -12.3% | -8.5% | -12.8% | -8.9% | -7.9% | -11.7% | -5.4% |
| inverted plunger | 19.4% | 1.0% | 2.3% | -9.0% | -8.1% | 9.6% | -10.6% | -10.3% | -3.4% | -3.1% | -2.8% |
| column | 2.4% | 4.7% | -1.2% | -6.6% | -5.6% | -5.6% | -4.3% | -10.9% | -1.9% | -2.2% | 0.2% |
| hourglass | 8.0% | 4.1% | -0.9% | -5.6% | -10.8% | -6.4% | -8.5% | -11.4% | -5.1% | -4.1% | 4.0% |

**Table S6** Frequency (counts) of shape series.

| From/to | pyramid | plunger | bell | lower diamond | middle diamond | upper diamond | inverted bell | inverted pyramid | inverted plunger | column | hourglass |
|---|---|---|---|---|---|---|---|---|---|---|---|
| 2 | 1184 | 212 | 775 | 1040 | 521 | 264 | 19 | 6 | 20 | 18 | 149 |
| 3 | 780 | 51 | 237 | 328 | 167 | 136 | 7 | 11 | 4 | 2 | 40 |
| 4 | 551 | 18 | 85 | 123 | 54 | 59 | 1 | - | 2 | 1 | 18 |
| 5 | 365 | 6 | 30 | 34 | 10 | 32 | - | - | 1 | - | 7 |
| 6 | 276 | 5 | 6 | 8 | 9 | 22 | - | - | 1 | - | 1 |
| 7 | 192 | 1 | 2 | 6 | 2 | 9 | - | - | - | - | 2 |
| 8 | 144 | - | 1 | - | - | 8 | - | 1 | - | - | - |
| 9 | 125 | 1 | - | - | - | 5 | - | - | - | - | - |
| 10 | 80 | - | - | - | - | 3 | - | - | - | - | 1 |
| 11 | 91 | 1 | - | - | - | 2 | - | - | - | - | 1 |
| 12 | 71 | - | - | - | - | - | - | - | - | - | - |
| 13 | 62 | - | - | - | - | 2 | - | - | - | - | - |
| 14 | 47 | - | - | - | - | 1 | - | - | - | - | - |
| 15 | 39 | - | - | - | - | - | - | - | - | - | - |
| 16 | 37 | - | - | - | - | - | - | - | - | - | - |
| 17 | 29 | - | - | - | - | - | - | - | - | - | - |
| 18 | 25 | - | - | - | - | - | - | - | - | - | - |
| 19 | 22 | - | - | - | - | - | - | - | - | - | - |
| 20 | 22 | - | - | - | - | - | - | - | - | - | - |
| 21 | 18 | - | - | - | - | - | - | - | - | - | - |
| 22 | 16 | - | - | - | - | - | - | - | - | - | - |
| 23 | 20 | - | - | - | - | - | - | - | - | - | - |
| 24 | 14 | - | - | - | - | - | - | - | - | - | - |
| 25 | 18 | - | - | - | - | - | - | - | - | - | - |
| 26 | 15 | - | - | - | - | - | - | - | - | - | - |
| 27 | 11 | - | - | - | - | - | - | - | - | - | - |
| 28 | 8 | - | - | - | - | - | - | - | - | - | - |
| 29 | 7 | - | - | - | - | - | - | - | - | - | - |
| 30 | 5 | - | - | - | - | - | - | - | - | - | - |
| 31 | 15 | - | - | - | - | - | - | - | - | - | - |
| 32 | 15 | - | - | - | - | - | - | - | - | - | - |
| 33 | 11 | - | - | - | - | - | - | - | - | - | - |
| 34 | 9 | - | - | - | - | - | - | - | - | - | - |
| 35 | 10 | - | - | - | - | - | - | - | - | - | - |
| 36 | 10 | - | - | - | - | - | - | - | - | - | - |
| 37 | 6 | - | - | - | - | - | - | - | - | - | - |
| 38 | 8 | - | - | - | - | - | - | - | - | - | - |
| 39 | 7 | - | - | - | - | - | - | - | - | - | - |
| 40 | 2 | - | - | - | - | - | - | - | - | - | - |
| 41 | 12 | - | - | - | - | - | - | - | - | - | - |
| 42 | 5 | - | - | - | - | - | - | - | - | - | - |
| 43 | 3 | - | - | - | - | - | - | - | - | - | - |
| 44 | 6 | - | - | - | - | - | - | - | - | - | - |
| 46 | 7 | - | - | - | - | - | - | - | - | - | - |
| 47 | 1 | - | - | - | - | - | - | - | - | - | - |
| 48 | 6 | - | - | - | - | - | - | - | - | - | - |
| 49 | 8 | - | - | - | - | - | - | - | - | - | - |
| 50 | 4 | - | - | - | - | - | - | - | - | - | - |
| 51 | 4 | - | - | - | - | - | - | - | - | - | - |
| 52 | 6 | - | - | - | - | - | - | - | - | - | - |
| 53 | 4 | - | - | - | - | - | - | - | - | - | - |
| 54 | 19 | - | - | - | - | - | - | - | - | - | - |

**Table S7** Percentage of population changes in shape series.

| From/to | pyramid | plunger | bell | lower diamond | middle diamond | upper diamond | inverted bell | inverted pyramid | inverted plunger | column | hourglass |
|---|---|---|---|---|---|---|---|---|---|---|---|
| 2 | 4.4% | 5.3% | -0.1% | -3.7% | -3.8% | -4.2% | -10.5% | -8.1% | -3.1% | -1.8% | 4.0% |
| | (±14.5) | (±16.3) | (±9.4) | (±8.1) | (±8.1) | (±9.0) | (±6.9) | (±6.5) | (±6.4) | (±6.2) | (±19.7) |
| 3 | 12.3% | 13.9% | 0.2% | -5.6% | -7.6% | -8.2% | -14.5% | -17.2% | -11.3% | -3.9% | 8.8% |
| | (±23.0) | (±26.4) | (±12.5) | (±10.8) | (±11.3) | (±11.9) | (±9.1) | (±11.4) | (±9.2) | (±3.9) | (±21.7) |
| 4 | 24.0% | 24.0% | -1.0% | -6.7% | -10.6% | -13.8% | -5.6% | - | -11.2% | -31.2% | 7.0% |
| | (±46.4) | (±63.6) | (±15.4) | (±12.4) | (±13.3) | (±12.5) | | | (±7.4) | | (±37.9) |
| 5 | 28.3% | 67.0% | 5.2% | -6.9% | -8.0% | -17.8% | - | - | -38.9% | - | 30.4% |
| | (±43.0) | (±95.4) | (±25.9) | (±11.3) | (±18.8) | (±13.7) | | | | | (±49.9) |
| 6 | 42.5% | 27.5% | -6.2% | 3.3% | -10.1% | -11.3% | - | - | 50.0% | - | -27.4% |
| | (±53.1) | (±41.5) | (±18.4) | (±8.6) | (±12.0) | (±17.5) | | | | | |
| 7 | 49.2% | 23.9% | 36.4% | 7.3% | -2.9% | -20.4% | - | - | - | - | -44.7% |
| | (±64.1) | | (±40.0) | (±13.4) | (±10.5) | (±14.7) | | | | | (±7.5) |
| 8 | 49.0% | - | 36.4% | - | - | -24.8% | - | -64.0% | - | - | - |
| | (±70.6) | | | | | (±14.4) | | | | | |
| 9 | 74.8% | 84.6% | - | - | - | -40.0% | - | - | - | - | - |
| | (±87.9) | | | | | (±20.2) | | | | | |
| 10 | 78.1% | - | - | - | - | -15.2% | - | - | - | - | -48.3% |
| | (±86.9) | | | | | (±12.4) | | | | | |
| 11 | 91.2% | 190.0% | - | - | - | -28.2% | - | - | - | - | -14.6% |
| | (±103.4) | | | | | (±14.0) | | | | | |
| 12 | 123.5% | - | - | - | - | - | - | - | - | - | - |
| | (±133.9) | | | | | | | | | | |
| 13 | 119.0% | - | - | - | - | -17.7% | - | - | - | - | - |
| | (±144.0) | | | | | (±2.9) | | | | | |
| 14 | 125.3% | - | - | - | - | -41.0% | - | - | - | - | - |
| | (±104.6) | | | | | | | | | | |
| 15 | 127.8% | - | - | - | - | - | - | - | - | - | - |
| | (±135.8) | | | | | | | | | | |
| 16 | 140.9% | - | - | - | - | - | - | - | - | - | - |
| | (±164.5) | | | | | | | | | | |
| 17 | 169.1% | - | - | - | - | - | - | - | - | - | - |
| | (±176.8) | | | | | | | | | | |
| 18 | 295.3% | - | - | - | - | - | - | - | - | - | - |
| | (±311.2) | | | | | | | | | | |
| 19 | 134.4% | - | - | - | - | - | - | - | - | - | - |
| | (±136.8) | | | | | | | | | | |
| 20 | 311.9% | - | - | - | - | - | - | - | - | - | - |
| | (±393.3) | | | | | | | | | | |
| 21 | 199.4% | - | - | - | - | - | - | - | - | - | - |
| | (±194.1) | | | | | | | | | | |
| 22 | 259.5% | - | - | - | - | - | - | - | - | - | - |
| | (±237.1) | | | | | | | | | | |
| 23 | 214.5% | - | - | - | - | - | - | - | - | - | - |
| | (±352.8) | | | | | | | | | | |
| 24 | 184.0% (±142.1) | - | - | - | - | - | - | - | - | - | - |
| 25 | 290.7% (±234.5) | - | - | - | - | - | - | - | - | - | - |
| 26 | 316.9% (±330.0) | - | - | - | - | - | - | - | - | - | - |
| 27 | 146.0% (±145.7) | - | - | - | - | - | - | - | - | - | - |
| 28 | 302.3% (±529.4) | - | - | - | - | - | - | - | - | - | - |
| 29 | 365.2% (±445.9) | - | - | - | - | - | - | - | - | - | - |
| 30 | 316.7% (±289.8) | - | - | - | - | - | - | - | - | - | - |
| 31 | 465.1% (±513.4) | - | - | - | - | - | - | - | - | - | - |
| 32 | 485.3% (±360.0) | - | - | - | - | - | - | - | - | - | - |
| 33 | 413.3% (±429.1) | - | - | - | - | - | - | - | - | - | - |
| 34 | 823.4% (±672.6) | - | - | - | - | - | - | - | - | - | - |
| 35 | 314.3% (±208.5) | - | - | - | - | - | - | - | - | - | - |
| 36 | 336.7% (±299.2) | - | - | - | - | - | - | - | - | - | - |
| 37 | 828.0% (±704.6) | - | - | - | - | - | - | - | - | - | - |
| 38 | 1475.6% (±2574.4) | - | - | - | - | - | - | - | - | - | - |
| 39 | 817.3% (±458.8) | - | - | - | - | - | - | - | - | - | - |
| 40 | 1109.5% (±488.0) | - | - | - | - | - | - | - | - | - | - |
| 41 | 1103.7% (±901.5) | - | - | - | - | - | - | - | - | - | - |
| 42 | 896.8% (±1257.1) | - | - | - | - | - | - | - | - | - | - |
| 43 | 981.6% (±546.9) | - | - | - | - | - | - | - | - | - | - |
| 44 | 736.2% (±721.2) | - | - | - | - | - | - | - | - | - | - |
| 46 | 1032.3% (±578.4) | - | - | - | - | - | - | - | - | - | - |
| 47 | 280.6% | - | - | - | - | - | - | - | - | - | - |
| 48 | 1177.4% (±1635.1) | - | - | - | - | - | - | - | - | - | - |
| 49 | 695.0% (±652.3) | - | - | - | - | - | - | - | - | - | - |
| 50 | 299.8% (±185.6) | - | - | - | - | - | - | - | - | - | - |
| 51 | 523.6% (±306.2) | - | - | - | - | - | - | - | - | - | - |
| 52 | 743.9% (±469.6) | - | - | - | - | - | - | - | - | - | - |
| 53 | 1309.7% (±697.1) | - | - | - | - | - | - | - | - | - | - |
| 54 | 449.7% (±385.8) | - | - | - | - | - | - | - | - | - | - |

***Table S8:*** *Classification rules based on the transitions between the n-th bucket and the (n+1)st bucket. An 'increase' is written as 1, a 'decrease' as -1 and a 0 stands for 'no change'. We assume the 'juvenile bucket' to be bucket number 1 and the 'senior bucket' to be bucket number 5.*

| Idealised transition sequence | Additional rules | Classification |
|---|---|---|
| -1, -1, 1, 1 | | Hourglass |
| -1, -1, 1, 0 | | Hourglass |
| -1, -1, 0, 1 | | Hourglass |
| 1, 1, 0, 0 | | Inverted Bell |
| 1, 0, 0, 0 | | Inverted Bell |
| 0, 1, 0, 0 | | Inverted Bell |
| 0, 0, -1, -1 | | Bell |
| 0, 0, -1, 0 | | Bell |
| 0, 0, 0, -1 | | Bell |
| 0, 0, 1, 1 | | Inverted Plunger |
| 0, 0, 1, 0 | | Inverted Plunger |
| 0, 0, 0, 1 | | Inverted Plunger |
| 1, 1, 1, 1 | | Inverted Pyramid |
| 1, 1, 1, 0 | | Inverted Pyramid |
| 1, 1, 0, 1 | | Inverted Pyramid |
| 1, 0, 1, 1 | | Inverted Pyramid |
| 0, 1, 1, 1 | | Inverted Pyramid |
| 0, 1, 1, 0 | | Inverted Pyramid |
| 0, 1, 0, 1 | | Inverted Pyramid |
| 1, 0, 1, 0 | | Inverted Pyramid |
| 1, 0, 0, 1 | | Inverted Pyramid |
| 1, -1, -1, -1 | | Lower Diamond |
| 1, -1, -1, 0 | | Lower Diamond |
| 1, -1, 0, -1 | | Lower Diamond |
| 1, -1, 0, 0 | | Lower Diamond |
| 1, 0, -1, -1 | | Lower Diamond |
| 1, 0, -1, 0 | | Lower Diamond |
| 1, 1, -1, -1 | | Diamond |
| 1, 1, -1, 0 | | Diamond |
| 1, 0, 0, -1 | | Diamond |
| 0, 1, -1, -1 | | Diamond |
| 0, 1, -1, 0 | | Diamond |
| -1, -1, 0, 0 | | Plunger |
| -1, 0, 0, 0 | | Plunger |
| 0, -1, 0, 0 | | Plunger |
| -1, -1, -1, -1 | | Pyramid |
| -1, -1, -1, 0 | | Pyramid |
| -1, -1, 0, -1 | | Pyramid |
| -1, 0, -1, -1 | | Pyramid |
| -1, 0, -1, 0 | | Pyramid |
| -1, 0, 0, -1 | | Pyramid |
| 0, -1, -1, -1 | | Pyramid |
| 0, -1, -1, 0 | | Pyramid |
| 0, -1, 0, -1 | | Pyramid |
| 1, 1, 1, -1 | | Upper Diamond |
| 1, 1, 0, -1 | | Upper Diamond |
| 1, 0, 1, -1 | | Upper Diamond |
| 0, 1, 1, -1 | | Upper Diamond |
| 0, 1, 0, -1 | | Upper Diamond |
| 0, 0, 1, -1 | | Upper Diamond |
| 0, 0, 0, 0 | | Column |
| -1, -1, -1, 1<br>0, -1, -1, 1 | B3 > B5<br>B3 <= B5 | Pyramid<br>Hourglass |
| 1, 1, -1, 1 | B3 > B5<br>B3 = B5<br>B3 < B5 | Diamond<br>Inverted Bell<br>Inverted Pyramid |
| 0, 0, -1, 1 | B3 > B5<br>B3 = B5<br>B3 < B5 | Bell<br>Column<br>Inverted Plunger |
| 0, 1, -1, 1 | B3 > B5<br>B3 = B5<br>B3 < B5 | Diamond<br>Inverted Bell<br>Inverted Pyramid |
| 1, 0, -1, 1 | B3 > B5<br>B3 = B5<br>B3 < B5 | Lower Diamond<br>Inverted Bell<br>Inverted Pyramid |
| 0, -1, 1, 0 | B2 > B4<br>B2 = B4<br>B2 < B4 | Plunger<br>Hourglass<br>Inverted Plunger |
| 0, -1, 0, 1 | B2 > B5<br>B2 = B5<br>B2 < B5 | Pyramid<br>Hourglass<br>Inverted Plunger |
| 1, -1, 1, 0 | B2 > B4<br>B2 = B4 | Lower Diamond<br>Inverted Bell |
| | B2 < B4 | Inverted Pyramid |
| 0, -1, 1, 1 | B2 > B4<br>B2 <= B4 | Hourglass<br>Inverted Plunger |
| -1, 0, 0, 1 | B2 > B5<br>B2 = B5<br>B2 < B5 | Plunger<br>Hourglass<br>Inverted Plunger |
| -1, 0, 1, 1<br>-1, 0, 1, 0 | B1 >= B4<br>B1 < B4 | Hourglass<br>Inverted Plunger |
| -1, 1, 1, 0 | B1 >= B4<br>B1 < B4 and B1 < B3<br>B1 < B4 and B1 >= B3 | Hourglass<br>Inverted Pyramid<br>Hourglass |
| -1, 1, -1, -1<br>-1, 1, -1, 0 | B1 > B3<br>B1 = B3<br>B1 < B3 | Pyramid<br>Bell<br>Diamond |
| -1, 1, 0, -1 | B1 > B3<br>B1 = B3<br>B1 < B3 | Pyramid<br>Bell<br>Upper Diamond |
| -1, 1, 0, 1<br>-1, 1, 1, 1 | B1 > B3<br>B1 = B3<br>B1 < B3 | Hourglass<br>Inverted Plunger<br>Inverted Pyramid |
| -1, 0, -1, 1 | B1 = B5 or B2 < B5<br>elseif B2 = B4<br>else | Hourglass<br>Plunger<br>Pyramid |
| -1, 1, 0, 0 | B1 > B3<br>B1 = B3<br>B1 < B3 | Plunger<br>Column<br>Inverted Bell |
| 1, -1, -1, 1 | B3 > B5<br>B3 = B5 and B1 > B3<br>B3 = B5 and B1 <= B3<br>B3 < B5 and B1 >= B3<br>B3 < B5 and B1 < B3 | Lower Diamond<br>Plunger<br>Lower Diamond<br>Hourglass<br>Inverted Pyramid |
| -1, 1, 1, -1 | B1 < B4<br>else<br>B1 > B3 and B3 > B5<br>else B3 < B5<br>else | Upper Diamond<br><br>Pyramid<br>Hourglass<br>Bell |
| -1, 0, 1, -1 | B1 <= B5<br>else<br>B1 > B4 | Upper Diamond<br><br>Pyramid |
| | B1 = B4<br>B1 < B4 | Bell<br>Hourglass |
| 1, -1, 1, -1 | B1 = B3 and B1 = B5<br>if B2 > B4<br>if B2 = B4<br>if B2 < B4<br>else if B1 > B3 and B3 > B5<br>else if B1 < B3 and B3 < B5<br>else if B1 > B3 and B3 < B5<br>B2 = B4<br>B2 > B4<br>B2 < B4<br>else<br>B2 = B4<br>B2 > B4<br>B2 < B4 | <br>Lower Diamond<br>Diamond<br>Upper Diamond<br>Pyramid<br>Inverted Pyramid<br><br>Hourglass<br>Lower Diamond<br>Upper Diamond<br><br>Hourglass<br>Lower Diamond<br>Upper Diamond |
| 0, -1, 1, -1 | B2 > B4 and B3 > B5<br>B2 > B4 and B3 = B5<br>B2 > B4 and B3 < B5<br>B2 = B4 and B3 < B5<br>B2 = B4 and B3 >= B5<br>B2 < B4 | Pyramid<br>Plunger<br>Hourglass<br>Hourglass<br>Bell<br>Upper Diamond |
| -1, -1, 1, -1 | B1 > B4 and B3 > B5<br>B1 > B4 and B3 = B5<br>B1 > B4 and B3 < B5<br>B1 = B4<br>B1 < B4 and B1 <= B5<br>B1 < B4 and B1 > B5 | Pyramid<br>Plunger<br>Hourglass<br>Bell<br>Upper Diamond<br>Hourglass |
| 1, -1, 1, 1 | B1 > B3<br>B1 < B3<br>B1 = B3 and B2 > B4<br>B1 = B3 and B2 = B4<br>B1 = B3 and B2 < B4 | Hourglass<br>Inverted Pyramid<br>Hourglass<br>Inverted Pyramid<br>Inverted Plunger |
| -1, 1, -1, 1 | B1 > B3 and B3 > B5<br>B1 > B3 and B3 < B5<br>B1 > B3 and B3 = B5<br>B1 = B3 and B3 = B5<br>B1 = B3 and B3 > B5<br>B1 = B3 and B3 < B5<br>B1 < B3 and B3 < B5 | Pyramid<br>Hourglass<br>Plunger<br>Column<br>Bell<br>Inverted Plunger<br>Inverted Pyramid |
| | B1 < B3 and B3 = B5 | Inverted Bell |
| | B1 < B3 and B3 > B5 | Diamond |
| 1, -1, 0, 1 | B1 = B3 and B2 <= B5 | Inverted Plunger |
| | B1 = B3 and B2 > B5 | Lower Diamond |
| | B1 > B3 and B2 < B5 | Hourglass |
| | B1 > B3 and B2 = B5 | Inverted Bell |
| | B1 > B3 and B2 > B5 | Lower Diamond |
| | B1 < B3 and B2 < B5 | Inverted Pyramid |
| | B1 < B3 and B2 = B5 | Inverted Bell |
| | B1 < B3 and B2 > B5 | Lower Diamond |